\documentclass[
prx,twocolumn,notitlepage,longbibliography,amsmath,amssymb,floats,superscriptaddress,nofootinbib,10pt]{revtex4-2}
\usepackage{graphicx,color}
\usepackage{amsmath}
\usepackage{amsfonts, amsbsy}
\usepackage{amssymb}
\usepackage{eqnarray}
\usepackage{bm}
\usepackage{blkarray}
\usepackage{mathtools}
\usepackage{psfrag}
\usepackage{multirow}
\usepackage{textcomp}
\usepackage{units}
\usepackage{lipsum}
\usepackage[title]{appendix}
\usepackage{soul}
\usepackage{titlesec}
\usepackage{times}
\usepackage{enumitem}
\usepackage{soul}
\usepackage[sort&compress]{natbib}
\usepackage[breaklinks,colorlinks = true,linkcolor = blue,urlcolor  = blue,citecolor = blue,anchorcolor = blue]{hyperref}
\usepackage{float}
\usepackage{orcidlink}

\def\beq{\begin{equation}}
\def\eeq{\end{equation}}

\newcommand{\be}{\begin{equation}}
\newcommand{\ee}{\end{equation}}
\newcommand{\ba}{\begin{align}}
\newcommand{\ea}{\end{align}}
\newcommand\eea{\end{eqnarray}}
\newcommand\bea{\begin{eqnarray}}
\newcommand{\bg}{\begin{gather}}
\newcommand{\eg}{\end{gather}}
\newcommand{\bseq}{\begin{subequations}}
\newcommand{\eseq}{\end{subequations}}

\newcommand{\mc}{\mathcal}

\newcommand{\DD}{\mathcal{D}}

\newcommand{\calL}{\mathcal{L}}

\newcommand{\olC}{\overline{C}}
\newcommand{\olK}{\overline{K}}

\newcommand{\tilT}{\widetilde{T}}
\newcommand{\tilcalL}{\widetilde{\mathcal{L}}}

\newcommand{\uu}{u} 
\newcommand{\torsion}{\phi} 
\newcommand{\bond}{\theta} 
\newcommand{\w}{\omega} 

\usepackage{xcolor}

\newcommand{\thint}{%
   \mathop{%
    \mathchoice
      {\!\vcenter{\hbox{\rule[0.6ex]{0.6em}{0.1ex}}}\kern-1.0em\int}
      {\!\vcenter{\hbox{\rule[0.6ex]{0.5em}{0.1ex}}}\kern-1.1em\int}
      {\!\vcenter{\hbox{\rule[0.5ex]{0.4em}{0.08ex}}}\kern-1.0em\int}
      {\!\vcenter{\hbox{\rule[0.4ex]{0.3em}{0.06ex}}}\kern-0.85em\int}%
  }%
}

\begin{document}
\title{Kinematic Foundations of Defect-Mediated Plasticity: A Gauge-Theoretic Framework}

\author{Kevin T. Grosvenor \orcidlink{}}
\email{kgrosvenor@nip.upd.edu.ph}
\affiliation{National Institute of Physics, University of the Philippines
Diliman, Quezon City 1101, Philippines}

\author{Mario Sol\'{i}s \orcidlink{0000-0001-6139-075X}}
\email{mario.solis-benites@pwr.edu.pl}
\affiliation{Institute of Theoretical Physics, Wroc\l{}aw University of Science and Technology, 50-370 Wroc\l{}aw, Poland}

\author{Piotr Sur\'owka \orcidlink{0000-0003-2204-9422}}
\email{piotr.surowka@pwr.edu.pl}
\affiliation{Institute of Theoretical Physics, Wroc\l{}aw University of Science and Technology, 50-370 Wroc\l{}aw, Poland}

\date{\today}

\begin{abstract}
Plastic deformation is widely regarded as an intrinsically dissipative phenomenon whose theoretical description relies heavily on phenomenological constitutive assumptions. We argue instead that defect-mediated plasticity possesses a universal, non-dissipative backbone: the kinematics and mobility constraints of defects are fixed by symmetry and conservation laws prior to dissipation. Starting from the spontaneous breaking of spacetime symmetries in a crystalline phase, we construct an effective field theory in which elasticity reorganizes into a coupled higher-rank tensor-vector gauge structure emerging directly from stress and defect conservation laws rather than being postulated microscopically. Dislocations, disclinations, and singular frame-geometric configurations appear as gauge sources whose coupled continuity equations and mobility constraints follow from generalized Gauss laws. The construction provides a symmetry-based organization of the kinematic sectors relevant to defect-mediated plasticity and a starting point for a subsequent dissipative theory of defect motion and stress relaxation.
\end{abstract}

\maketitle

\section{Introduction}

Plastic deformation is one of the most universal and technologically consequential phenomena in condensed matter physics
\cite{lubliner_plasticity_1990,hosford_fundamentals_2013,zaccone2023theory}.
In crystalline solids, irreversible deformation is mediated by the motion,
creation, annihilation, and interaction of defects. Its dynamics is commonly
described through constitutive flow rules, phenomenological defect mobilities,
and stress-relaxation rates. These ingredients are necessarily
material-dependent. They nevertheless operate subject to a more universal
kinematic structure: defect charges obey compatibility and continuity
relations, different defect species can terminate on or transform into one
another only in restricted ways, and the motion of individual defects is
constrained by conservation laws.

The purpose of this work is to identify this kinematic structure independently
of a particular dissipative closure. We do not construct here a complete theory
of irreversible plastic flow, which would additionally require defect
mobilities, nucleation and proliferation mechanisms, entropy production, and
stress relaxation. Instead, we determine the conservative field content, gauge
redundancies, and defect conservation laws that constrain such a theory. In this
sense, the construction below provides a kinematic foundation for
defect-mediated plasticity rather than a complete theory of plastic deformation.

Our central conjecture is that the appropriate long-wavelength starting point for defect-mediated plasticity is precisely such a symmetry-constrained, nondissipative theory of coupled elastic and geometric defects. According to this conjecture, irreversible plastic flow should be formulated as a dissipative constitutive completion of this kinematic sector, rather than by
introducing defect kinetics and flow rules independently from the outset. The
present work does not establish the full dissipative theory or prove the uniqueness of this organizing principle. It provides evidence for the conjecture by constructing the underlying coupled gauge structure and showing that it reproduces defect continuity relations, termination rules, and mobility constraints without phenomenological assumptions about defect motion.

A long-standing approach to crystalline defects is to formulate them
geometrically and gauge-theoretically; see
Refs.~\cite{Sahoo2006GaugeI,ValsakumarSahoo2006GaugeII} for reviews.
Beginning with the work of Kondo, Bilby, and others
\cite{Kondo1953FoundationsYielding,KondoEtAl1968NonholonomicGeometryPlasticity,
BilbyBulloughSmith1955ContinuousDistributions,Kroner1985ContinuumTheoryDefects},
dislocations and disclinations were associated with torsion and curvature of an
effective material manifold. This insight led to gauge theories in which
elastic degrees of freedom couple to fields associated with local translations
and rotations, while defect densities are constrained by geometric identities.

Two broad directions emerged. Yang--Mills-type theories with gauge groups such
as $SO(3)\ltimes T(3)$ introduce gauge fields for dislocations and
disclinations
\cite{KadicEdelen1983GaugeTheoryDislocationsDisclinations,
ValsakumarSahoo1988,Lazar2002ElastoplasticTheoryDislocationsTorsion,
Lazar2003NonsingularEdgeDislocation,Lazar2008-LAZTGT-3}.
Gravity-inspired approaches instead interpret defects as torsion and curvature
of a Riemann--Cartan material geometry
\cite{KatanaevVolovich1992TheoryDefects3DGravity,Katanaev2004GeometricTheoryDefects},
with plastic distortion represented by a non-integrable geometry generated by
multivalued coordinate transformations
\cite{kleinert_multivalued_2008}. These formulations have clarified the
geometric meaning of crystalline defects and produced, among other results,
regularized defect cores and the correct far-field stresses. At the same time,
different gauge realizations can encode the same defect geometry, leaving some
freedom in the choice and physical interpretation of the gauge variables.

A complementary development is provided by elasticity dualities and, more
recently, fracton--elasticity duality
\cite{
kleinert_duality_1982,
kleinert_double_1983,
beekman_dual_3d_2017,
beekman_dual_2d_2017,
pretko_fracton_elasticity_2018,kumar_symmetry-enforced_2019,
pretko_crystal--fracton_2019,
radzihovsky_fractons_2020,
radzihovsky_quantum_smectic_2020,
gromov_cosserat_fractons_2020,nguyen_fracton-elasticity_2020,
surowka_quasicrystal_2021,gaa_fracton-elasticity_2021,
manoj_fractonic_2021,hirono_qi_2022,caddeo_emergent_2022,
beekman_crystal_gravity_2022,PhysRevE.105.024602,
tsaloukidis_fracton-elasticity_2024,TsaloukidisSurowka2024,
tsaloukidis_nonequilibrium_2025,glodkowski_quadrupole_2025,
matus_defects_2026}.
These constructions solve the stress-conservation equations of elasticity in
terms of higher-rank gauge potentials. Dislocations and disclinations then
appear as tensor gauge charges, while their continuity equations and restricted
mobility follow from generalized Gauss laws. The subdimensional character of
defect motion is therefore a kinematic consequence of elastic constraints
encoded in higher-rank gauge symmetry.

Existing fracton--elasticity constructions predominantly focus on the
translational elastic sector. Even extensions incorporating curved backgrounds
or metric fluctuations generally retain an effectively Riemannian geometry, in
which independent torsional and frame-rotational degrees of freedom are absent
or represented only implicitly by singular configurations. Geometric defect
theories, by contrast, incorporate torsion and curvature directly, but their
relation to the stress-based tensor gauge variables is not usually obtained by
dualizing a common action. The relation between translational,
frame-rotational, and torsional sectors therefore remains insufficiently
resolved when several types of defects coexist and can act as sources for one
another.

In this work, we address this kinematic gap in a linearized theory in
$2+1$ dimensions. Starting from the symmetry-breaking structure of a
crystalline phase, we retain the displacement and bond-angle fields together
with torsional-frame and connection degrees of freedom. Dualizing the resulting
conservation equations yields two rank-two tensor gauge fields coupled to two
vector $U(1)$ sectors. The tensor and vector fields are not independent:
Stueckelberg-like couplings organize them into a single constrained gauge
structure. In this way, the gauge variables arise as potentials solving the
conservation laws of the elastic--geometric theory rather than being chosen by
analogy with a predetermined Yang--Mills or gravitational model.

The construction leads to three main results. First, gauge invariance produces
coupled defect continuity equations. These equations determine which charges
are conserved independently, which defect currents can act as sources for
other charges, and which termination and conversion processes are
kinematically allowed. In particular, dislocation charge need not be conserved
separately: a bond-angle current can act as its source, expressing the
geometric statement that dislocations may terminate on disclinations. An
analogous hierarchy occurs in the additional frame-geometric sector, while the
two vector charges are conserved only in combination.

Second, the theory separates smooth propagating modes from singular defect
sectors. The symmetry-broken phase contains gapless phonons, whereas the
independent rotational and frame-geometric fluctuations generically acquire a
gap through their coupling to the elastic moduli. Although these massive smooth
modes can be integrated out at long wavelengths, their multivalued
configurations survive as topological sources in the dual theory. The
low-energy theory therefore retains the defect content of geometric sectors
whose ordinary fluctuations are gapped.

Third, the tensor Gauss constraint provides a common description of
dislocation glide and climb. In the absence of vacancies and interstitials, it
implies conservation of a projected dipole moment and restricts a dislocation
to motion parallel to its Burgers vector. Coupling to point-defect density
modifies this constraint, allowing the projected moment to change and thereby
permitting climb. The physical statement that climb requires the exchange of
vacancies or interstitials is well known; the present construction identifies
how this process modifies the Gauss constraint within the enlarged
tensor--vector gauge theory.

The torsional sources appearing in this formulation should not be interpreted
as a new microscopic species of crystalline defect. They describe singular
configurations of the antisymmetric, or frame, part of the effective material
geometry. The vielbein contains information not only about metric deformation
but also about the orientation and non-integrability of the local material
frame. Its smooth frame fluctuations generally become massive in the
crystalline phase, but their singular configurations cannot in general be
eliminated together with the smooth modes. They survive as effective gauge
charges describing collective non-integrable configurations of conventional
crystalline defects. In this sense, torsional-frame defects complement the
metric defects arising in curved formulations of fracton--elasticity duality:
the two sectors encode singularities of the antisymmetric and symmetric parts
of the effective material geometry, respectively.

The resulting framework determines conservative defect interactions,
continuity relations, mobility constraints, and selection rules for defect
reactions. It does not determine dissipative mobilities, plastic flow curves,
defect proliferation rates, or stress-relaxation mechanisms. These require a
subsequent dissipative extension. The present theory instead supplies the
symmetry-constrained kinematic sector that such constitutive dynamics must
respect, thereby separating universal geometric constraints from
material-specific transport.

This logic is not unique to elasticity. Recent progress in magnetohydrodynamics (MHD) has shown that identifying the correct higher-form gauge structure of ideal conducting fluids is essential before dissipative effects (such as resistivity and viscosity) can be incorporated consistently within a hydrodynamic expansion (see e.g. \cite{iqbal_lectures_2025}). In that context, the systematic construction of dissipative MHD relied not on phenomenological assumptions, but on a precise characterization of conserved currents, generalized symmetries, and gauge redundancies of the underlying ideal theory. Plasticity requires, and we show admits, an analogous reorganization.

The remainder of this paper is organized as follows. In Sec.~II we introduce
the linearized elastic and frame-geometric fields and their symmetry-allowed
couplings. In Sec.~III we dualize the resulting conservation equations and
derive the coupled tensor--vector gauge structure. In Sec.~IV we admit singular
field configurations, identify the corresponding defect charges and currents,
and derive their continuity equations and mobility constraints, including the
glide constraint and its relaxation by vacancies and interstitials. In
Sec.~V we compare the resulting framework with geometric defect theory,
fracton--elasticity duality, and continuum approaches to plasticity, and
discuss the dissipative extension required for a theory of irreversible
plastic flow. Technical derivations are collected in the appendices.

\section{Geometric starting point}

Elasticity theory admits a natural interpretation as the low-energy effective description of a
spontaneously broken spacetime symmetry. In the absence of a crystalline structure, the system
is invariant under independent Poincar\'e transformations acting on the material and laboratory
coordinates. The emergence of a solid phase corresponds to a spontaneous breaking of this
symmetry down to a diagonal subgroup, locking material and spatial frames. In this sense,
elastic media may be viewed as ordered phases of a theory with underlying
$ISO(1,d)_{\text{material}}\times ISO(1,d)_{\text{spatial}}$ symmetry broken to $ISO(1,d)_{\text{diag}}$, where $d$ corresponds to the dimension of the space.

In the conventional Landau picture, elasticity arises by expanding the effective action around
this symmetry-broken state to quadratic order in the Goldstone fields associated with broken
translations and rotations. This procedure yields the familiar elastic energy functional expressed
in terms of displacement and rotational degrees of freedom, and is sufficient for describing
small deformations about a homogeneous reference configuration. However, such an expansion
implicitly assumes a fixed background geometry and obscures the geometric origin of defects
and their associated conservation laws.

A systematic field-theoretic formulation of elasticity proceeds instead by first introducing a
background geometric structure and subsequently allowing for its crystallization. In this
approach, the elastic medium is described as a geometric phase defined on a spacetime manifold,
with the ordered crystalline state selecting a preferred local frame. Elastic deformations,
defects, and their interactions are then understood as fluctuations and singularities of this
geometric structure, rather than as properties of an a priori fixed reference configuration.

Since plastic deformation and defect dynamics generically involve non-integrable distortions,
the appropriate background geometry must allow for torsion. In torsionful geometries, the
fundamental geometric variables are the vielbein $e^{a}{}_{\mu}$, which encodes local translations
and defines the metric structure, and the spin connection $\omega^{a}{}_{b\mu}$, which governs
local rotations.

In the weak–field regime relevant for small elastic and plastic distortions, we expand the vielbein and spin connection around a flat reference configuration. This procedure parallels weak–field expansions in first-order formulations of gravity, where geometric fluctuations are treated perturbatively about a trivial background (see, e.g., \cite{karananas_particle_2015}). We parametrize the fluctuation as \[
e^m{}_i = \delta^m_i + h^m{}_i , \qquad |h|\ll 1 ,
\]
\[
\omega^{mn}{}_i , \qquad |\omega^{mn}{}_i| \ll 1 ,
\]
where \(m,n\) denote local Lorentz indices and \(i,j\) label spatial coordinates. The background configuration fixes a reference identification between material and spatial frames, while the fluctuations \(h^m{}_i\) encode departures from this ordered state. The geometric formulation involves an independent spin connection \(\omega^m{}_{n i}\), which governs local rotations of the material frame. In the weak--field expansion, we similarly expand the spin connection around a trivial background.

Lowering the index with the background Kronecker delta, \(h_{ij}=\delta_{mj}\,h^m{}_i\), it is natural to decompose the vielbein fluctuation into its symmetric and antisymmetric parts,
\begin{align}
s_{ij} &= \frac12\bigl(h_{ij}+h_{ji}\bigr),\\
a_{ij} &= \frac12\bigl(h_{ij}-h_{ji}\bigr).
\end{align}
The symmetric tensor \(s_{ij}\) corresponds to the conventional elastic strain, while the antisymmetric component \(a_{ij}\) describes local rotational distortions of the material frame. This decomposition makes explicit the separation between metric and rotational degrees of freedom already implicit in the geometric formulation, and will be convenient for organizing the linearized dynamics. 

Expanding the action to second order in fluctuations around the flat background, the quadratic theory naturally organizes itself according to the underlying geometric fields. In particular, the spin connection \(\omega\), the symmetric strain \(s_{ij}\), and the antisymmetric mode \(a_{ij}\) each contribute independent quadratic terms, together with symmetry--allowed couplings between them. The quadratic action may be written schematically as
\begin{align}
S_2 &= S_2(\omega,\omega)
+ S_2(s,s)
+ S_2(a,a) \nonumber \\
&\quad + S_2(\omega,s)
+ S_2(\omega,a)
+ S_2(s,a)\, .
\end{align}
This structure reflects the fact that translations and rotations are not independent in a torsionful geometry: elastic strains, local rotations, and connection degrees of freedom are dynamically intertwined already at the linearized level. The mixing terms encode the geometric constraints relating curvature, torsion, and frame deformations, and will play an important role in determining the spectrum and kinematics of defect excitations.

\subsection{Total action and field content}

Elasticity can be understood as the low–energy effective theory of a phase in
which spatial translations and rotations are spontaneously broken by the
formation of a crystal. In the symmetric phase the system is invariant under
independent translations of the material and laboratory coordinates, while in
the solid phase these symmetries are locked to a diagonal subgroup. The true
Goldstone modes associated with this breaking are the displacement fields
$\uu_i$, describing fluctuations of atoms around equilibrium lattice
positions.

The orientational scalar $\bond$, encoding local bond–angle rotations, is not
an independent Goldstone mode. Instead, it behaves as a massive
Stückelberg-like field: its fluctuations acquire a gap once the crystal forms,
so bond–angle distortions represent massive rotational excitations of the
lattice frame rather than additional gapless modes.
Following the effective–field–theory viewpoint, the ordered phase should be
constructed by coupling these Goldstone modes to background geometric
variables and then expanding around a homogeneous reference configuration.
In this picture, torsion and connection degrees of freedom represent geometric
distortions that can exist independently of the phonons. Importantly, once the
crystal forms, these geometric modes generically acquire a gap through their
coupling to the elastic moduli, while the phonon modes remain gapless. This is
analogous to the Higgs or Stückelberg mechanism familiar from gauge theories,
where the elastic tensor plays the role of a mass matrix for geometric
fluctuations.

In the geometric description the vielbein fluctuation can be decomposed into
symmetric and antisymmetric parts. In the present work we set the symmetric
part to zero and focus on torsional geometries with vanishing metric fluctuations. This restriction isolates the torsional sector relevant for defect geometry while leaving the full geometric formulation available for generalizations.
Accordingly, the relevant geometric degrees of freedom are torsion and the
spin–connection–like field $\w_i$, which capture non-integrable distortions
associated with defects.

The construction is general and can be formulated in $3$ spatial
dimensions. However, for simplicity we specialize to $2+1$ dimensions, where rotational distortions reduce to scalars and the theory admits a dual gauge-field
formulation in which defects appear as sources of conserved currents.

We consider the real-time path integral
\begin{equation}
Z=\frac{1}{Z_0}\!\int\!\DD\uu\,\DD a\,\DD\bond\,\DD\w_i\;
e^{iS[\uu,a,\bond,\w]},
\end{equation}
with
\begin{equation}
S=\int dt\,d^2x\,\calL[\uu,a,\bond,\w].
\end{equation}

Spontaneous breaking of spatial translations produces the gapless Goldstone
modes $\uu_i$. Bond-angle fluctuations $\bond$ are not independent Goldstones;
rather, they behave as massive Stückelberg-like fields associated with local
rotational redundancy of the lattice frame.

We also introduce geometric variables describing local frame deformations: a
torsional scalar $\torsion$ and a connection $\w_i$. In two dimension, the antisymmetric torsion part and the spin-connection
\be
a_{ij}= \epsilon_{ij} \torsion\,, \quad \omega^i\,_{k \ell} = \epsilon_{k\ell}\w_i\,.
\ee
At the present stage we work in the defect-free sector, so all fields are taken to be smooth and
single-valued and the geometry is integrable. In particular, we restrict to
vanishing curvature and, correspondingly, the symmetric (metric) part of the
vielbein fluctuation is set to zero. Torsion and curvature become nontrivial
only once singular configurations are admitted, which will be discussed later.
\begin{equation}
S=S_{\rm el}+S_{\rm geom}.
\end{equation}

\paragraph{Elastic sector.}
Local rotational invariance requires the combinations
\begin{equation}
\gamma_{ij}=\partial_i\uu_j-\epsilon_{ij}(\bond+\torsion),\qquad
\partial_i\bond+\w_i .
\end{equation}
The elastic action reads
\begin{align}
S_{\rm el}=\!\int\!dt\,d^2x\,
\Big[
&\tfrac12 \dot{\uu}_i\dot{\uu}_i
-\tfrac12 C_{ijk\ell}\gamma_{ij}\gamma_{k\ell} \notag\\
&+\tfrac12 \dot{\bond}^2
-\tfrac12\beta(\partial_i\bond+\w_i)^2
\Big].
\end{align}
In a crystalline medium spatial translations and rotations are spontaneously broken by the elastic medium, while time translations remain unbroken.

\paragraph{Geometric sector.}
The geometric fields possess independent quadratic dynamics,
\begin{align}
S_{\rm geom}=\!\int\!dt\,d^2x\,
\Big[
&\tfrac12 \dot{\torsion}^2
-\tfrac12 v^2(\partial_i\torsion)^2 \notag\\
&+\tfrac12 \dot{\w}_i\dot{\w}_i
-\tfrac12 K_{ijk\ell}\partial_i\w_j\partial_k\w_\ell
\Big].
\end{align}
In our construction the spatial manifold is allowed to possess torsion, 
but we do not formulate a full Poincaré gauge theory in which torsion 
arises as the field strength of an independent translational gauge field. 
Instead, the crystalline medium fixes the local spatial frame and thereby 
Higgses the translational gauge redundancy. As a consequence, only the 
physical propagating modes that survive this gauge fixing are retained, 
leading to an effective description in which the remaining torsional 
degree of freedom appears as a reduced gauge–invariant mode.

\subsection{Equations of motion}
\label{sec:EOM}

We now derive the equations of motion for the elastic–geometric theory defined in Eqs.~(7) and (8). These equations describe the coupled dynamics of phonons, bond--angle fluctuations, and
geometric torsional degrees of freedom prior to dualization.

The action is written in terms of the displacement field $u_i$,
torsional scalar $\phi$, bond--angle field $\theta$, and
spin--connection--like field $\omega_i$,
\begin{equation}
S=\int dt\, d^2x\, \mathcal{L}[u_i,\phi,\theta,\omega_i],
\end{equation}
with rotationally invariant strain tensor
\begin{equation}
\gamma_{ij}=\partial_i u_j-\epsilon_{ij}(\theta+\phi).
\end{equation}
Varying the action with respect to each field yields a set of
coupled conservation laws and wave equations encoding the
interplay between elastic and geometric distortions.

Variation with respect to the displacement field $u_j$ gives
\begin{equation}\label{eq:momentum_eq}
\ddot{u}_j-\partial_i\!\left(C_{ijk\ell}\gamma_{k\ell}\right)=0 .
\end{equation}
Variation with respect to $\phi$ yields
\begin{equation}
\ddot{\phi}-v^2\partial_i \partial^i \phi-\epsilon_{ij}T_{ij}=0,
\end{equation}
showing that torsional distortions are sourced by the
antisymmetric part of the stress tensor.
Analogously, the equation of motion for $\theta$ is
\begin{equation}
\ddot{\theta}-\beta\,\partial_i(\partial_i\theta+\omega_i)
-\epsilon_{ij}T_{ij}=0.
\end{equation}
Finally, variation with respect to $\omega_j$ gives
\begin{equation}\label{eq:omega_eom}
\ddot{\omega}_j-\partial_i(K_{ijk\ell}\partial_k\omega_\ell)
+\beta(\partial_j\theta+\omega_j)=0.
\end{equation}
Eqs.~(\ref{eq:momentum_eq})--(\ref{eq:omega_eom})
describe the coupled propagation of gapless phonons together
with mostly massive rotational and geometric excitations. The
structure of these equations already hints at the emergence of
gauge redundancy and conservation laws that become manifest
in the dual gauge-field formulation in analogy with Cosserat elasticity. However, since only one equation is of the conservation form the identification of the gauge variables requires some care.

\section{Dual gauge formulation}
\label{sec:dual}

The equations of motion derived in the previous section
already contain the essential ingredients of a gauge-theoretic
description. In particular, the displacement equation takes the
form of a conservation law,
\begin{equation}
\partial_t P_j-\partial_i T_{ij}=0 ,
\label{eq:stress_conservation_repeat}
\end{equation}
expressing local conservation of linear momentum. As in the
case of ordinary elasticity, such conservation laws can be
solved identically by introducing gauge potentials whose field
strengths reproduce the conserved currents. The purpose of
this section is to construct the corresponding dual
gauge-field formulation for the full elastic–geometric theory.

\subsection{Stress conservation and tensor gauge fields}

Equation~(\ref{eq:momentum_eq}) may be written
compactly as
\begin{equation}
\partial_\mu T^\mu{}_j = 0 ,
\end{equation}
where $T^t{}_j=P_j$ and $T^i{}_j=-T_{ij}$. The conservation
law can be solved identically by introducing a vector-valued
gauge field $A_{\mu j}$ such that the stress tensor is written
as the curl of a gauge potential,
\begin{equation}
T^\mu{}_j=\epsilon^{\mu\nu\rho}\partial_\nu A_{\rho j}.
\label{eq:T_dual_def}
\end{equation}
This representation automatically satisfies the conservation
law and introduces the gauge redundancy
\begin{equation}
A_{\mu j}\rightarrow A_{\mu j}+\partial_\mu \alpha_j .
\end{equation}

Separating temporal and spatial components,
\begin{equation}
A_{t j}\equiv \Phi_j, \qquad A_{i j}\equiv A_{ij},
\end{equation}
the momentum density and stress tensor become
\begin{align}
P_j &= \epsilon_{ik}\partial_i A_{kj}, \\
T_{ij} &= \epsilon_{ik}(\partial_t A_{kj}-\partial_k\Phi_j).
\end{align}
It is convenient to introduce generalized electric and magnetic
fields,
\begin{align}
B_j &= \epsilon_{ik}\partial_i A_{kj},\\
E_{ij} &= \epsilon_{ik}(\partial_t A_{kj}-\partial_k\Phi_j),
\end{align}
so that $P_j=B_j$ and $T_{ij}=\epsilon_{ik}\epsilon_{j\ell}
E_{k\ell}$.

At this stage the displacement field has been traded for a
rank-two tensor gauge field.

\subsection{Dualization strategy}

The remaining equations of motion couple the antisymmetric
stress $\epsilon_{ij}T_{ij}$ to the scalar fields $\phi$ and
$\theta$. Using Eq.~(\ref{eq:T_dual_def}) one finds
\begin{equation}
\epsilon_{ij}T_{ij}=\partial_t A_{ii}-\partial_i\Phi_i .
\label{eq:trace_stress}
\end{equation}
The dualization proceeds by introducing auxiliary fields
enforcing the definitions of the stress and torque currents
and integrating out the original elastic and geometric fields.
Operationally, this amounts to a sequence of
Hubbard–Stratonovich transformations that trade the
second-order elastic action for a first-order action written
entirely in terms of gauge-field strengths.

\subsection{Dualization of the torsional scalar}
\label{sec:dual_phi}

We next dualize the torsional scalar $\phi$.  
Using the tensor-gauge representation of the stress tensor,
the antisymmetric stress entering the scalar equation of motion
may be written as
\begin{equation}
\epsilon_{ij}T_{ij}=\partial_t A_{ii}-\partial_i\Phi_i .
\end{equation}
The equation of motion of $\phi$ therefore becomes
\begin{equation}
\ddot{\phi}-v^2\nabla^2\phi
-\left(\partial_t A_{ii}-\partial_i\Phi_i\right)=0 .
\label{eq:phi_eom_dual}
\end{equation}

Equation~(\ref{eq:phi_eom_dual}) can be written as a conservation
law by defining the current
\begin{equation}
\tilde L_t=\partial_t\phi-A_{ii},
\qquad
\tilde L_i=v^2\partial_i\phi-\Phi_i .
\end{equation}
The scalar equation of motion then takes the form
\begin{equation}
\partial_t \tilde L_t-\partial_i\tilde L_i=0 ,
\end{equation}
which is solved identically by introducing a $U(1)$ gauge field
$\tilde a_\mu$ such that
\begin{equation}
\tilde L_\mu=\epsilon_{\mu\nu\rho}\partial^\nu \tilde a^\rho .
\end{equation}
The torsional scalar is thus traded for the electric and magnetic
fields of a vector gauge field.

Unlike the displacement sector, the current $\tilde L_\mu$
contains the tensor gauge fields explicitly. As a consequence,
the tensor gauge symmetry is not manifest after dualization.
Gauge invariance is restored by enlarging the gauge
transformations so that the combinations
\begin{equation}
\tilde L_t=\partial_t\phi-A_{ii},\qquad
\tilde L_i=v^2\partial_i\phi-\Phi_i
\end{equation}
remain invariant. This requires a simultaneous transformation
of the tensor and vector gauge fields, leading to a coupled
gauge structure in which the $U(1)$ and tensor sectors are no
longer independent.

In terms of the electric and magnetic fields of $\tilde a_\mu$,
\begin{equation}
\tilde b=\epsilon_{ij}\partial_i\tilde a_j,
\qquad
\tilde e_i=\partial_t\tilde a_i-\partial_i\tilde a_t ,
\end{equation}
the torsional sector contributes
\begin{equation}
\mathcal L_{\phi}
= -\frac12(\tilde b + A_{ii})^2
+ \frac{1}{2v^2}(\epsilon_{ij}\tilde e_j + \Phi_i)^2 .
\label{eq:Lphi_dual}
\end{equation}

This result demonstrates an important feature of the theory:
the torsional scalar does not produce an independent gauge
sector, but instead couples directly to the trace of the tensor
gauge field. This mixing will play a central role in the full
dual gauge theory.

\subsection{Dualization of the bond--angle field}
\label{sec:dual_theta}

We now dualize the bond--angle field $\theta$.  
Using the tensor gauge representation of the stress tensor,
the equation of motion for $\theta$,
\begin{equation}
\ddot{\theta}-\beta\,\partial_i(\partial_i\theta+\omega_i)
-\epsilon_{ij}T_{ij}=0 ,
\end{equation}
can be rewritten in the same form as the torsional scalar
equation.

Defining the current
\begin{equation}
L_t=\partial_t\theta-A_{ii},
\qquad
L_i=\beta(\partial_i\theta+\omega_i)-\Phi_i ,
\end{equation}
the equation of motion becomes the conservation law
\begin{equation}
\partial_t L_t-\partial_i L_i=0 .
\end{equation}
This continuity equation is solved by introducing a second
$U(1)$ gauge field $a_\mu$,
\begin{equation}
L_\mu=\epsilon_{\mu\nu\rho}\partial^\nu a^\rho ,
\end{equation}
with electric and magnetic fields
\begin{equation}
b=\epsilon_{ij}\partial_i a_j,
\qquad
e_i=\partial_t a_i-\partial_i a_t .
\end{equation}

As in the torsional sector, the current $L_\mu$ depends
explicitly on the tensor gauge fields $(\Phi_i,A_{ij})$.
Gauge invariance is therefore restored only after enlarging the
combined gauge transformations of the tensor and $U(1)$
sectors. The two vector gauge fields are thus not independent,
but are intertwined with the tensor gauge symmetry.

The bond--angle sector contributes
\begin{equation}
\mathcal L_{\theta}
= -\frac12(b + A_{ii})^2
+ \frac{1}{2\beta}(\epsilon_{ij}e_j + \Phi_i)^2 .
\label{eq:Ltheta_dual}
\end{equation}

The close similarity between
Eqs.~(\ref{eq:Lphi_dual}) and (\ref{eq:Ltheta_dual})
reflects the parallel roles of torsional and bond--angle
distortions. Together they generate two coupled $U(1)$ gauge
fields interacting with the tensor gauge sector.

\subsection{Dualization of the spin connection}
\label{sec:dual_omega}

We now complete the dualization by resolving the spin–connection
sector. The equation of motion for $\omega_j$ reads
\begin{equation}
\ddot{\omega}_j-\partial_i(K_{ijk\ell}\partial_k\omega_\ell)
+\beta(\partial_j\theta+\omega_j)=0 .
\end{equation}
Using the dual representation of the bond–angle sector derived
above, this equation can be rewritten as a conservation law.

After substituting the dual expressions for the bond–angle and
torsional sectors, the equation of motion can be written in the
form
\begin{equation}
\partial_t \tilde P_j-\partial_i \tilde T_{ij}=0 ,
\end{equation}
where the conserved momentum and stress are
\begin{align}
\tilde P_j &= \partial_t\omega_j
+ \epsilon_{ij}(\tilde a_i-a_i),\\
\tilde T_{ij} &= K_{ijk\ell}\partial_k\omega_\ell
+ \epsilon_{ij}(\tilde a_t-a_t).
\end{align}
Thus the spin connection generates a second conserved
stress tensor. This conservation law is solved identically by introducing a
second rank-two tensor gauge field $\tilde A_{\mu j}$,
\begin{equation}
\tilde T^\mu{}_j
=\epsilon^{\mu\nu\rho}\partial_\nu \tilde A_{\rho j}.
\end{equation}
Separating temporal and spatial components,
\begin{equation}
\tilde A_{tj}\equiv \tilde\Phi_j ,
\qquad
\tilde A_{ij}\equiv \tilde A_{ij},
\end{equation}
the conserved quantities become
\begin{align}
\tilde P_j &= \epsilon_{ik}\partial_i\tilde A_{kj},\\
\tilde T_{ij} &= \epsilon_{ik}
(\partial_t\tilde A_{kj}-\partial_k\tilde\Phi_j).
\end{align}

In terms of the electric and magnetic fields of the second
tensor gauge field,
\begin{equation}
\tilde B_j=\epsilon_{ik}\partial_i\tilde A_{kj},
\qquad
\tilde E_{ij}=\epsilon_{ik}
(\partial_t\tilde A_{kj}-\partial_k\tilde\Phi_j),
\end{equation}
the geometric sector contributes
\begin{equation}
\begin{split}
\mathcal L_{\omega}
= & -\frac12
\left(\tilde B_j-\epsilon_{jk}(\tilde a_k-a_k)\right)^2
\\
& + \frac12 \tilde C_{ijk\ell}
\left(\tilde E_{k\ell}
-\epsilon_{k\ell}(\tilde a_t-a_t)\right)^2 .
\end{split}
\label{eq:Lomega_dual}
\end{equation}

Collecting the results of the previous subsections, the dual
description is expressed in terms of two rank-two tensor gauge
fields $(A_{\mu j},\tilde A_{\mu j})$ together with two
interacting $U(1)$ gauge fields $(a_\mu,\tilde a_\mu)$. These
sectors are not independent: the elastic and geometric
constraints generate Stückelberg-like couplings that tie the
tensor and vector gauge fields into a single coupled gauge
structure.

It is important to verify that the dual formulation carries the
same number of propagating degrees of freedom as the original
elastic–geometric theory.  The microscopic description
contains four fields $(u_i,\phi,\theta,\omega_i)$, corresponding
to six real dynamical components in $2+1$ dimensions. At the moment we ignore the fact that certain components of the geometric fields can be further set to zero without changing the physics. This would obscure the dual gauge structure.

In the dual description the field content is enlarged to two
rank-two tensor gauge fields $(A_{\mu j},\tilde A_{\mu j})$ and
two $U(1)$ gauge fields $(a_\mu,\tilde a_\mu)$.  At first sight
this appears to introduce many additional variables.  However,
the dual theory possesses an extended set of gauge
transformations that remove the redundant components.  The
tensor gauge fields admit vector gauge parameters, while the
two $U(1)$ sectors carry scalar gauge redundancies.  Moreover,
the Stückelberg-like couplings linking the tensor and vector
sectors eliminate the would-be independent gauge symmetries
associated with separate transformations of the $U(1)$ fields,
leaving only the diagonal combination as a true gauge
invariance.

The dual Lagrangian is invariant under a coupled set of gauge
transformations with independent parameters
$\alpha_k$, $\tilde\alpha_k$, $\lambda$, and $\tilde\lambda$.
These transformations act as
\begin{subequations}\label{eq:gauge_combined}
\begin{align}
\tilde\Phi_k 
&\to \tilde\Phi_k + \partial_t \tilde\alpha_k ,
&
\tilde A_{ik} 
&\to \tilde A_{ik} + \partial_i \tilde\alpha_k ,
\label{eq:gauge_combined_a}
\\[4pt]
\Phi_k 
&\to \Phi_k + \partial_t \alpha_k ,
&
A_{ik} 
&\to A_{ik} + \partial_i \alpha_k ,
\label{eq:gauge_combined_b}
\\[4pt]
a_t 
&\to a_t + \partial_t \lambda ,
&
\tilde a_t 
&\to \tilde a_t + \partial_t \lambda ,
\label{eq:gauge_combined_c}
\\[4pt]
a_i 
&\to a_i + \partial_i \lambda + \epsilon_{ij}\alpha_j ,
&
\tilde a_i 
&\to \tilde a_i + \partial_i \lambda + \epsilon_{ij}\alpha_j .
\label{eq:gauge_combined_d}
\end{align}
\end{subequations}

These transformations ensure the invariance of the dual
Lagrangian and remove the redundant gauge components of the
tensor and vector potentials. The dual formulation therefore
contains the same number of propagating degrees of freedom as
the original elastic–geometric theory. These transformations ensure the invariance of the dual Lagrangian and remove redundant gauge components of the tensor and vector potentials. After accounting for the resulting gauge redundancies and the St\"uckelberg-type constraints that eliminate the relative $U(1)$ symmetry between the two vector sectors, the Dirac counting yields six propagating degrees of freedom in $2+1$ dimensions, matching the six dynamical modes of the original elastic--geometric theory $(u_i,\phi,\theta,\omega_i)$.

After accounting for these gauge redundancies, the number of
physical propagating modes coincides with that of the original
elastic–geometric theory.  The dual formulation therefore
provides a faithful and complete rewriting of the low-energy
degrees of freedom in terms of coupled tensor and vector gauge
fields.

It is important to clarify the role of the massive geometric modes in the
low-energy description. In the symmetry-broken phase, torsional and
rotational fluctuations generically acquire a gap through their coupling
to the elastic moduli, and may therefore be integrated out when
constructing a long-wavelength effective theory. However, integrating
out massive smooth fluctuations does not eliminate the corresponding
topological sectors. Singular configurations of these fields, encoding
defects, remain as nontrivial gauge charges in the dual formulation.
Their continuity equations and termination rules are protected by gauge
redundancy and therefore persist in the infrared theory.

As a result, the low-energy effective action retains a memory of the
massive geometric sector through induced couplings between the remaining
gauge fields and through the structure of defect interactions. In this
sense, the systematic dualization procedure separates propagating
massive modes from topological defect degrees of freedom, allowing one
to integrate out the former without discarding the latter.

\section{Defects as Gauge Charges}

The defect sectors relevant to plasticity enter once singular field configurations are admitted. We therefore allow the displacement, bond-angle, torsional, and frame fields 
to contain multivalued components. These singular parts encode topological 
defects. 

Separating each field into regular and singular contributions and integrating 
out the smooth sector, the action reduces to a 
minimal coupling between gauge potentials and conserved sources,
\begin{equation}
\begin{split}
S_{\mathrm{def}}
= \int dt\, d^2x \,
\Big(
& \rho_k \Phi_k + J_{ik} A_{ik} 
+ j_\mu a^\mu + \tilde j_\mu \tilde a^\mu \\
& + \tilde\rho_k \tilde\Phi_k 
+ \tilde J_{ik} \tilde A_{ik}
\Big).
\end{split}
\end{equation}
All defect information is therefore encoded in gauge charges and currents. The construction of the defect currents from singular field configurations and their explicit expressions are given in 
Appendix~\ref{app:source_defects}.

The tensor charges $\rho_k$ arise from noncommuting derivatives of the 
singular displacement field and correspond to dislocations. The vector 
charges $j_\mu$ originate from singular bond-angle configurations and 
describe disclinations. The remaining sources encode torsional and 
frame-rotation defects. In this formulation, defects appear as gauge 
charges required by non-integrable geometry rather than as externally 
imposed singularities.

Gauge invariance of the dual theory enforces the continuity equations 
(Appendix \ref{app:continuity_eqs}),
\begin{subequations} \label{eq:ContEqs}
\begin{align}
\partial_t \rho_k + \partial_i J_{ik} - \epsilon_{ik}j_i &=0\,, \label{eq:DislocationContEq}   \\
\partial_t \tilde{j}_t + \partial_t j_t + \partial_i \tilde{j}_i + \partial_i j_i &= 0 \,,  \\
\partial_t \tilde{\rho}_k + \partial_i\tilde{J}_{ik} - \epsilon_{ik} \tilde{j}_i &= 0 \,.
\end{align}
\end{subequations}
These relations are not postulated but follow directly from gauge redundancy. The resulting continuity relations generalize the classical kinematic equations of defect dynamics introduced by Kossecka and de Wit \cite{KosseckaDeWit1977_Kinematics}. Equation \eqref{eq:DislocationContEq}, for example, expresses the fact that dislocations are not
independently conserved: they may terminate on disclinations, whose current $j_i$ acts as a source.

The first equation shows that dislocation charge is not conserved 
independently: it can be sourced by bond-angle current. 
Geometrically, this expresses the statement that dislocations may 
terminate on disclinations. In contrast, the vector charges obey 
ordinary continuity equations, reflecting the topological conservation 
of disclination and torsional charge in the absence of higher sources. 
The final equation reveals an analogous hierarchy in the geometric sector, 
where frame defects are sourced by torsional currents.

These coupled continuity equations encode the allowed defect reactions. 
They determine which defects may nucleate, annihilate, or transmute, 
and which processes are forbidden. In particular, isolated tensor charges 
inherit mobility constraints from the Gauss laws of the tensor gauge 
sector, leading to subdimensional or restricted motion characteristic of 
fracton-elasticity duality. 

The coupled gauge charges encode the conservative kinematics underlying defect-mediated plasticity. The Maxwell-like equations summarized in 
Appendix \ref{app:maxwell_eqs} determine defect interactions and mobility restrictions at the 
level of the ideal theory, prior to the introduction of dissipative effects.


\subsection{Gauge Constraints and the Glide Principle}

A concrete physical consequence of the gauge structure is the emergence of the glide constraint for dislocations. In the present framework, dislocations appear as gauge charges of a higher-rank tensor gauge field. In particular, the dislocation density is $\rho_i$ and its current tensor is $J_{ij}$. Therefore, setting to zero the disclination current $j_i$, the continuity equation \eqref{eq:DislocationContEq} becomes
\begin{equation}
    \partial_t \rho_{j} + \partial_i J_{ij}=0.
\label{eq:DefectContinuity}
\end{equation}
The structure of the gauge theory further implies that, in the absence of vacancy or interstitial degrees of freedom, the defect current is traceless,
\begin{equation}
J_{ii} = 0 .
\label{eq:TracelessCurrent}
\end{equation}
To extract the kinematic content of these relations, consider the scalar moment
\begin{equation}
M \equiv \int d^2x\, x_i \rho_i .
\label{eq:ProjectedMoment}
\end{equation}
Taking a time derivative and using \eqref{eq:DefectContinuity},
\begin{equation}
\frac{dM}{dt}
= \int d^2x\, x_j \dot{\rho}_j = - \int d^2x\, x_j \partial_i J_{ij}.
\end{equation}
Integrating by parts, assuming vanishing boundary flux, and using \eqref{eq:TracelessCurrent},
\begin{equation}
\frac{dM}{dt}
= \int d^2x\, J_{ii} = 0.
\end{equation}
Thus, the projected dipole moment $M$ is conserved as a direct consequence of gauge invariance.

The tracelessness of $J_{ij}$ in the absence of interstitials can be understood from different perspectives. The most obvious of these is that for a single dislocation with a constant Burgers vector $\mathbf{b}$ located at position $\mathbf{X} (t)$,
\begin{equation} \label{eq:burger}
\rho_i (x,t) = \epsilon_{ij} b_j \delta^{(2)}( \mathbf{x} - \mathbf{X}(t)),
\end{equation}
so that
\begin{equation} \label{eq:projmom}
M = \epsilon_{ij} X_i b_j.
\end{equation}
Conservation of $M$ implies
\begin{equation} \label{eq:glide}
\epsilon_{ij} \dot{X}_i b_j = 0,
\end{equation}
which is precisely the glide constraint in 2D: the velocity of a dislocation must be parallel to its Burgers vector. Transverse motion (climb) would change the conserved moment and is therefore forbidden unless additional degrees of freedom are introduced that relax the Gauss constraint.

It is important to emphasize that the theory does not enforce conservation of the full dipole tensor $\int d^2x\, x_i \rho_j$, which would imply complete immobility. Rather, only the projected moment \eqref{eq:projmom} is conserved. The resulting mobility is therefore subdimensional: dislocations may glide freely parallel to their Burgers vector, while climb requires coupling to vacancy or interstitial fields that modify the constraint structure.

In this way, the glide principle emerges directly from gauge redundancy and Gauss constraints, without imposing phenomenological mobility rules. It is a manifestation of the underlying symmetry-determined structure of the defect sector.

A complementary perspective is emphasized in \cite{Zaanen2006}: dislocations ``do not carry volume.'' Thus, they cannot couple to compressional stress (i.e., $J_{ii} = 0$) in the absence of interstitials or vacancies.

The glide constraint is derived in \cite{Zaanen2006} within the framework of
continuum elasticity. Physically, climb
requires the emission or absorption of vacancies or interstitials and is
therefore incompatible with the mass conservation of linear
elasticity.

Our derivation differs conceptually in two respects.
First, the constraint emerges here directly from the Gauss law of the translational gauge sector.
The glide principle therefore appears as a consequence of gauge
redundancy rather than as an additional kinematic assumption.
Second, the Gauss constraint implies the conservation of the projected
dipole moment $M$, leading immediately to the condition
\eqref{eq:glide} for a single dislocation.
This makes explicit the connection between dislocation glide and the
dipole conservation laws that appear in tensor gauge theories.


\subsection{Vacancies and the Restoration of Climb}

Vacancies and interstitials are naturally incorporated in our framework by modifying the Gauss constraint by an
additional gradient source term. This relaxes the conservation of the projected dipole moment $M$ and restores climb motion.

Let $n_v$ denote the vacancy/interstitial density field. Henceforth, we will use ``vacancy'' to refer to either vacancies or interstitials. This modifies the continuity equation \eqref{eq:DefectContinuity} to
\begin{equation}
\dot{\rho}_j + \partial_i J_{ij}
= - \partial_j \dot{n}_v,
\label{eq:ModifiedContinuity}
\end{equation}
where $\partial^j n_v$ represents the contribution of vacancy density to the
defect sector. Physically, this expresses the fact that point defects act as
sources or sinks of Burgers vector when dislocations absorb or emit
vacancies. In our framework, $n_v$ is not introduced by hand, but, comparing with equation \eqref{eq:DislocationContEq}, it is clear that $n_v$ is related to the source $j_i$ for the $U(1)$ gauge field $a_{i}$ via
\begin{equation}
    j_i = - \epsilon_{ij} \partial_j \dot{n}_v
\end{equation}
The trace of the current is now no longer constrained to vanish. Instead,
\begin{equation}
J_{ii} = - \dot{n}_v ,
\label{eq:TraceWithVacancies}
\end{equation}
so that compressional motion of dislocations is accompanied by time-dependent
vacancy density.

Now, the time evolution of the projected moment is
\begin{equation}
\frac{dM}{dt}
= \int d^2x\, J_{ii}
= - \int d^2x\, \dot{n}_v .
\end{equation}
Thus the projected dipole moment is no longer conserved; its variation is
controlled by vacancy dynamics. In particular, for a single dislocation with Burgers vector $\mathbf{b}$ at position $\mathbf{X} (t)$,
\begin{equation}
    \epsilon_{ij} \dot{X}_i b_j \neq 0,
\end{equation}
whenever vacancy density changes in time. Motion transverse to the Burgers vector (climb) is therefore permitted precisely when vacancies or
interstitials are exchanged with the defect core.

In this way climb appears not as a violation of gauge symmetry, but as a controlled relaxation of the Gauss constraint through coupling to additional fields. Glide is symmetry-protected in the pure defect sector, while climb is mediated by vacancy dynamics. This provides a systematic and
field-theoretic realization of the classical statement that dislocation climb requires mass transport.


\subsection{Relation to defect-mediated melting and hydrodynamic approaches}

The modern understanding of defect-driven phase transitions in two
dimensions originates from the work of Nelson and Halperin
\cite{NelsonHalperin1979}, where melting was shown to proceed via
dislocation and disclination unbinding. In that framework,
topological defects are introduced as singular configurations of the
elastic field, and their fugacities control the renormalization of
elastic constants. The dynamical extension developed in
\cite{ZippeliusHalperinNelson1980} incorporated mobile defects into
hydrodynamics, demonstrating how shear rigidity is lost through
defect proliferation and how defect motion relaxes stress.

More recently, hydrodynamic effective theories such as
\cite{ArmasEtAl2023} have organized plasticity directly at the level
of symmetry and transport, introducing additional relaxation channels
associated with plastic deformations. In these formulations,
defect dynamics enters implicitly through phenomenological relaxation
rates, allowing interpolation between solid-like and liquid-like
response in electronic crystals and related systems.

The present construction differs conceptually from both programs.
Rather than introducing defects as singular configurations whose
statistical mechanics or kinetics must be specified independently,
we derive defect currents as gauge charges required by non-integrable
geometry. Their continuity equations follow from gauge redundancy,
and admissible termination and transmutation processes are fixed by
Gauss constraints. In this sense, the defect-mediated melting
scenario of \cite{NelsonHalperin1979,ZippeliusHalperinNelson1980}
and the hydrodynamic plasticity framework of \cite{ArmasEtAl2023}
can be viewed as dynamical realizations built on top of a deeper
symmetry-determined kinematic structure. The dual tensor gauge
formulation identifies this structure explicitly.

\section{Discussion}

In this work, we constructed a nondissipative theory of the coupled defect
kinematics underlying defect-mediated plasticity. Starting from the symmetry-breaking structure of a crystalline phase, we derived a coupled tensor--vector gauge formulation in which defect continuity equations, mobility constraints, and admissible termination processes follow from gauge redundancy. The theory does not describe irreversible plastic flow itself: defect mobilities, nucleation and proliferation, entropy production, and stress relaxation require an additional dissipative constitutive extension. Its purpose is to identify the symmetry-constrained kinematic sector that such an extension must respect.

Gauge-theoretic approaches to defects have historically bifurcated into Yang–Mills-type and gravity-type constructions, with no consensus on which symmetry principle is fundamental. In many cases, gauge potentials were introduced by analogy rather than derived from first principles, leading to ambiguity regarding their physical interpretation and dynamical status. Our construction resolves this ambiguity by deriving the gauge structure directly from symmetry breaking and conservation laws. The rank-two tensor gauge fields arise as dual potentials solving stress conservation identically, while additional U(1) gauge sectors emerge from torsional and bond-angle currents. The resulting theory is a coupled tensor–vector gauge system in which the gauge symmetries are not postulated but forced by kinematics. In this sense, the appropriate gauge structure of elasticity is neither purely Yang–Mills nor purely gravitational; it is a higher-rank tensor gauge theory dictated by elastic conservation laws. This removes the longstanding ambiguity concerning which gauge symmetry underlies defect physics and provides a unified, symmetry-based derivation of the dynamical variables.

In classical defect geometry, the Bianchi identities encode kinematic statements such as the impossibility of disclination termination within the bulk and the fact that dislocations may end on disclinations. While geometrically elegant, these relations are typically imposed at the level of differential identities rather than derived from dynamical principles. In the present framework, defect continuity equations arise directly from gauge invariance of the dual theory. Once singular field configurations are admitted, defects appear as gauge charges coupled minimally to tensor and vector potentials. The continuity equations governing these charges follow from the invariance of the action under gauge transformations. Consequently, termination rules and defect transmutation processes are not additional assumptions but unavoidable consequences of Gauss constraints. Dislocation charge is not independently conserved but may be sourced by bond-angle currents, reflecting the geometric hierarchy between torsion and curvature. The structure of defect kinematics is therefore encoded in gauge redundancy itself, elevating classical geometric identities to consequences of symmetry.

Geometric approaches to defect theory identify dislocations with torsion and disclinations with curvature of a material manifold. However, the dynamical interplay between these geometric quantities has remained obscure, particularly regarding how rotational distortions, torsion, and elastic strains interact at the field-theoretic level. In our construction, torsional and bond-angle fields dualize to U(1) gauge sectors that couple directly to the trace components of the tensor gauge fields through Stueckelberg-like terms. This reveals that torsion, curvature, and rotational distortions are not independent topological sectors but components of a single constrained gauge system. The coupling structure generates a mass hierarchy analogous to a Higgs mechanism: gapless phonons emerge from broken translations, while geometric modes associated with frame rotations and torsion acquire a gap. Thus, the geometric variables that previously appeared as independent ingredients are unified dynamically within a single gauge-theoretic framework. The theory clarifies how curvature- and torsion-based descriptions fit into a coherent effective field theory of elasticity and plasticity.

While fracton–elasticity dualities have established that
conventional crystalline elasticity maps to higher-rank tensor
gauge theories, these constructions typically assume a fixed
background torsion in which only translational Goldstone
modes are retained and curvature or torsion enter implicitly
through singular configurations. In contrast, the present
framework derives the coupled tensor–vector gauge structure
directly from the spontaneous breaking of spacetime
symmetries, treating translational, rotational, and torsional
sectors on equal footing from the outset. The spin connection
and torsional degrees of freedom are not auxiliary or imposed
by analogy, but arise systematically in the symmetry-based
construction and participate dynamically in the dual theory. As a consequence, defect motion is not introduced as an
external kinetic assumption but follows from the gauge
structure itself. Dislocation currents appear as tensor gauge
charges whose continuity equations and mobility constraints
are fixed by Gauss laws, while their coupling to rotational
and torsional sectors determines the allowed transmutation
and termination processes. This unified treatment allows one
to describe not only static defect kinematics, but the coherent
propagation and interaction of dislocations within a
symmetry-determined field theory. In this sense, the theory
extends fracton–elasticity duality beyond a correspondence
between strain and tensor gauge fields, providing a dynamical
framework in which defect motion and geometric backreaction
are encoded within a single coupled gauge system.

A persistent structural issue in nonlinear field dislocation mechanics concerns the absence of a symmetry-determined foundation fixing the gauge content of the theory. In the crystal plasticity framework of \cite{Acharya2001}, closure of the governing equations requires eliminating the null-space freedom of the curl operator through additional structural assumptions, ensuring uniqueness of the elastic distortion while maintaining consistency with nonlinear elasticity. The finite-deformation formulation developed in \cite{Acharya2004} derives thermodynamically consistent driving forces for dislocation motion and nucleation, but the geometric sector remains fixed through constitutive structure rather than emerging from symmetry. More recently, the action-based reformulation in \cite{Acharya2022,acharya_dual_2023} constructs variational principles corresponding to nonlinear dislocation dynamics. Continuum formulations of dislocation dynamics, such as those developed in \cite{Acharya2001,Acharya2004,Acharya2022,acharya_dual_2023}, provide a kinematically complete description in terms of physically observable fields, with compatibility and transport relations imposed as part of the governing equations. The present approach does not modify this physical content, but instead reformulates it within a tensor gauge framework, at the linearized level and incorporating Cosserat-type (microrotational) degrees of freedom, in which defect densities and compatibility conditions are encoded geometrically in terms of torsion and curvature. This provides an alternative, symmetry-based organization of the same underlying structure, and establishes a framework amenable to effective field theory methods, where the gauge structure accommodates higher-form symmetries, enabling systematic coarse-graining and controlled scale separation within a symmetry-constrained expansion. Defect continuity equations, termination rules, and mobility constraints therefore follow from gauge redundancy itself, eliminating the residual ambiguity concerning which fields are fundamental and which are artifacts of representation.

Having identified this conservative backbone, the natural next step is to introduce dissipation in a controlled manner. Just as viscous hydrodynamics is constructed as a derivative expansion around ideal Euler flow, dissipative plasticity may be organized as a systematic deformation of the tensor–vector gauge theory. The gauge structure identified here constrains the admissible transport coefficients, defect mobilities, and relaxation channels that can appear once time-reversal symmetry is broken. Rather than postulating phenomenological flow rules, one can derive the structure of plastic transport directly from symmetry and effective field theory principles. In this way, plasticity joins a growing class of condensed matter systems whose long-wavelength structure is dictated not by phenomenology, but by gauge redundancy and symmetry.

\section*{Acknowledgements}

We are grateful to Aleksander G\l\'{o}dkowski, Francisco Pe\~na-Ben\'itez, and Lazaros Tsaloukidis for discussions. We thank Amit Acharya for comments on the manuscript and pointing out an error in our characterization of glide in an earlier version of this work. KTG acknowledges the Oﬃce of the Chancellor of the University of the Philippines Diliman, through the Oﬃce of the Vice Chancellor for Research and Development, for funding support
through the Outright Research Grant (Grant No. 262608 ORG). MS and PS were supported in part by the Polish National Science Centre (NCN) Sonata Bis grant 2019/34/E/ST3/00405.


\appendix

\begin{widetext}

\section{Hubbard-Stratonovich transformation and duality}
\label{app:A}

The full partition function of our model is 
\be
Z = \frac{1}{Z_0} \int \DD\uu \DD\torsion \DD \bond \DD\w_i \, e^{iS[\uu, \torsion, \bond , \w]}\,.
\ee
with
\be
S[\uu, \torsion, \bond , \w] = \int dt d^2 x\, \calL[\uu, \torsion, \bond , \w] \,.
\ee
The Lagrangian consists of four contributions associated with the
displacement field, scalar torsion, bond angle, and spin connection:
\be
\calL[\uu,\torsion,\bond,\w]
=
\calL[\uu]
+\calL[\torsion]
+\calL[\bond]
+\calL[\w]\,.
\ee
These four contributions are given explicitly by:
\bea
\calL[\uu_i] &=& \tfrac{1}{2} \dot{\uu}_i \dot{\uu}_i - \tfrac{1}{2} C_{ijk \ell} \gamma_{ij} \gamma_{k \ell}\,, \\
\calL[\torsion] &=& \tfrac{1}{2} \dot{\torsion}^2 - \tfrac{1}{2} v^2 \partial_i \torsion \, \partial_i \torsion \,, \\
\calL[\bond] &=& \tfrac{1}{2} \dot{\bond}^2 - \tfrac{1}{2} \beta ( \partial_i \bond + \omega_i ) ( \partial_i \bond + \w_i )\,, \\
\calL[\w_i] &=& \tfrac{1}{2} \dot{\w}_i \dot{\w}_i - \tfrac{1}{2} K_{ijk \ell} \, \partial_i \w_j \, \partial_k \w_{\ell}\,,
\eea
where 
\be
\gamma_{ij}=\partial_i\uu_j-\epsilon_{ij}(\bond+\torsion)\,.
\ee
We consider the elastic and geometric tensor constants have the following symmetries
\bea\label{eq:symmetries_C_K}
C_{ijk\ell} &=& C_{k\ell ij}\,,  \\
K_{ijk\ell} = K_{k\ell ij}\,, &\quad & K_{ijk\ell} =-K_{jik\ell} = - K_{ij \ell k}\,.
\eea
The constants $\beta$ and $v$ are physical constants for the bond angle and torsion (geometric) part.

The Hubbard–Stratonovich (HS) representation of the path integral is
\be
Z_{HS} = \frac{1}{Z_0} \int \DD[\mc{T}_{ij}] \, e^{iS_{HS}}\,,
\ee
where the integral measure is the product of all the HS fields $\DD[\mc{T}_{ij}] \equiv \DD P_k \DD T_{ij} \DD L_t \DD L_i \DD \tilde{P}_k \DD \tilde{T}_{ij} \DD \tilde{L}_t \DD \tilde{L}_i $. The Hubbard-Stratonovich action is
\be
S_{HS} = S^* + S_{\text{source}}\,.
\ee
In order to write these two pieces in terms of HS fields, it is convenient to use some gauge dual fields. Therefore, let us introduce the set of gauge transformation
\be
T_{tj}\equiv P_j = \dot{\uu}_j \,,, \quad  T_{ij} = C_{ijk \ell} \gamma_{k \ell}\,.
\ee
If $C$ is invertible, its inverse is denoted by
\be
\olC_{ijmn} C_{mnk \ell} = \delta_{ik} \delta_{j \ell}\,,
\ee
the inverse $\olC$ follows the same pattern of symmetries as $C$ on the eq. \eqref{eq:symmetries_C_K}. Then,
\be
\gamma_{ij} = \olC_{ijk \ell} T_{k \ell}.
\ee

We can resolve the equations of motions, as we did on the main text, for $P_k$ and $T_{ik}$ in terms of gauge fields $A_{tk} \equiv \Phi_k$ and $A_{jk}$ as follows
\be \label{eq:Treso}
P_k = \epsilon_{ij} \partial_i A_{jk}\,, \quad T_{ik} = \epsilon_{ij} ( \partial_t A_{jk} - \partial_j \Phi_k )\,.
\ee
Following the steps explained on section \ref{sec:dual}, in case for the torsion part, we use the relations
\be\label{eq:b}
\tilde{L}_t = \partial_t \torsion - A_{ii}\,, \quad \tilde{L}_i = v^2 \partial_i \torsion - \Phi_i\,,
\ee
after solved the equation of motion, the gauge fields $(\tilde{a}_t\,, \tilde{a}_i)$ are related to the HS fields as
\be 
\tilde{L}_t = \epsilon_{ij} \partial_i \tilde{a}_j \equiv \tilde{b}\,, \quad  \tilde{L}_i = \epsilon_{ij} ( \partial_t \tilde{a}_j - \partial_j \tilde{a}_t )\equiv \epsilon_{ij}\tilde{e}_j\,.
\ee
The next case, for the bond angle $\bond$, we have
\be 
L_t = \partial_t \bond - A_{ii} \epsilon_{ij} \partial_i a_j\equiv b\,, \quad L_i = \beta ( \partial_i \bond + \omega_i ) - \Phi_i = \epsilon_{ij} ( \partial_t a_j - \partial_j a_t )\equiv \epsilon_{ij}e_j\,.
\ee
The last dual piece comes from the spin connection, we define
\be
\tilde{P}_j = \partial_t \w_j + \epsilon_{ij} \bigl( \tilde{a}_i - a_i \bigr)\,, \quad \tilT_{ij} = K_{ijk \ell} \partial_k \w_{\ell} + \epsilon_{ij} \bigl( \tilde{a}_t - a_t  \bigr) - v^2 \torsion \delta_{ij}\,.
\ee
Providing we can invert $K$, its inverse is denoted $\olK$ and defined by
\be
\olK_{ijmn} K_{mnk \ell} = \delta_{ik} \delta_{j \ell}\,,
\ee
also the tensor constant $\olK$ follows the same symmetry as $K$ given in the equation \eqref{eq:symmetries_C_K}.
Therefore, the HS fields and the gauge dual are
\be
\tilde{P}_k =  \epsilon_{ij} \partial_i \tilde{A}_{jk}\,, \quad \tilT_{ik} = \epsilon_{ij} (\partial_t \tilde{A}_{jk} - \partial_j \tilde{\Phi}_k)\,.
\ee

Finally, the HS action is
\bea\label{eqs:HS_action_full}
S^* &=& \frac{1}{2}\int dt d^2x\,  \left\lbrace - P_kP_k + \tilde{C}_{ijk\ell} T_{ik}T_{j\ell} - (\tilde{L}_t + A_{ii})^2 + \frac{1}{v^2} (\tilde{L}_i + \Phi_i)^2  -(L_t + A_{ii})^2 + \frac{1}{\beta} (L_i + \Phi_i)^2 \right. \nonumber \\
&&\left. -  \bigl[\tilde{P}_{k} - \epsilon_{ik} \bigl( \tilde{a}_i - a_i \bigr) \bigr] \bigl[ \tilde{P}_k - \epsilon_{\ell k} \bigl( \tilde{a}_\ell - a_\ell \bigr)) \bigr] + \olK_{ijk \ell} \bigl[ \tilT_{k \ell}- \epsilon_{k \ell} \bigl( \tilde{a}_t - a_t \bigr) \bigr] \bigl[ \tilT_{ij} - \epsilon_{ij} \bigl(  \tilde{a}_t - a_t  \bigr) \bigr]  \right\rbrace \,,
\eea
and the source action part is
\bea\label{eq:source_lagrangian}
S_{\text{source}} &=&\int dt d^2x\, \left \lbrace P_i  \partial_t \uu_i - T_{ij}\gamma_{ij} + (\tilde{L}_t +A_{ii})\partial_t \torsion  - (\tilde{L}_i + \Phi_i) \partial_i \torsion  +  (L_t + A_{ii}) \partial_t \theta - (L_i + \Phi_i) \partial_i  \bond \right.\nonumber \\
&& \left. + \left( \tilde{P}_j - \epsilon_{ij}(\tilde{a}_i - a_i)\right) \partial_t \w_j  - \left( \tilT_{ij} - \epsilon_{ij}(\tilde{a}_t - a_t) \right) \partial_i\w_j \right\rbrace\,.
\eea
The HS action written in this way is instructive and convenient for the next calculation, and it enables us to appreciate how the theory is coupled between the terms. 

\section{Sources and defects}
\label{app:source_defects}

We now write the source action using Eq.~\eqref{eq:source_lagrangian}. 
After expressing the fields in terms of the gauge variables and decomposing them 
into smooth and singular parts, we integrate out the smooth sector and obtain
\bea\label{eq.source_action_1}
S_{\text{source}} &=& \int dt d^2 x\, \left[ P_i \partial_t \uu_i^{(s)} - T_{ij}\gamma_{ij}^{(s)} + (\tilde{L}_t +A_{ii})\partial_t \torsion^{(s)} - (\tilde{L}_i + \Phi_i) \partial_i \torsion^{(s)} +  (L_t + A_{ii}) \partial_t \theta^{(s)} - (L_i + \Phi_i) \partial_i \bond^{(s)}  \right. \nonumber \\
&& \left. + \left( \tilde{P}_j - \epsilon_{ij}(\tilde{a}_i - a_i)\right) \partial_t \w_j^{(s)} - \left( \tilT_{ij} - \epsilon_{ij}(\tilde{a}_t - a_t) \right) \partial_i\w_j^{(s)} \right]\,.
\eea
Integrating by parts, we obtain
\be\label{eq.source_action_2}
S_{\text{source}}  = \int dt d^2 x\, \left(\rho_k \Phi_k + J_{ik} A_{ik}  +   \tilde{j}_t \tilde{a}_t +  \tilde{j}_j \tilde{a}_j + j_t a_t +  j_j a_j  + \tilde{\rho}_k \tilde{\Phi}_k + \tilde{J}_{ik} \tilde{A}_{ik}  \right)\,,
\ee
where the currents and charges are
\bea
\rho_k = \epsilon_{ij} \partial_i \partial_j u_k^{(s)}\,, &\quad& J_{ik} = \epsilon_{ij} (\partial_j \partial_t - \partial_t \partial_j   ) \uu_k^{(s)} \,, \label{eq:rhok} \\
j_t = \epsilon_{ij} \partial_i \partial_j \bond^{(s)} - \epsilon_{ij}\partial_i\w_j^{(s)} \,, &\quad& j_j = \epsilon_{ij}(\partial_t \partial_i -\partial_i \partial_t) \bond^{(s)} + \epsilon_{jk} \partial_t \w_k^{(s)} \,, \\
\tilde{j}_t = \epsilon_{ij} \partial_i \partial_j \torsion^{(s)} + \epsilon_{ij}\partial_i\w_j^{(s)}\,, &\quad& \tilde{j}_j = \epsilon_{ij}(\partial_t \partial_i -\partial_i \partial_t) \torsion^{(s)} - \epsilon_{jk} \partial_t \w_k^{(s)}\,, \\
\tilde{\rho}_k = \epsilon_{ij} \partial_i \partial_j \w_k^{(s)}\,, &\quad& \tilde{J}_{ik} = \epsilon_{ij} (\partial_j \partial_t - \partial_t \partial_j) \w_k^{(s)} \,.
\eea
The spin connection contributes with opposite signs to the two vector charge and current densities. Consequently, it cancels from their sums: 
\bea j_t+\tilde{j}_t &=& \epsilon_{ij}\partial_i\partial_j \left(\bond^{(s)}+\torsion^{(s)}\right), \\ j_j+\tilde{j}_j &=& \epsilon_{ij} \left( \partial_t\partial_i-\partial_i\partial_t \right) \left(\bond^{(s)}+\torsion^{(s)}\right). 
\eea 
Thus, the two vector sectors encode complementary contributions from the crystalline bond-angle field and the torsional-frame field, while the spin-connection contribution transfers charge between them. This structure anticipates the coupled continuity equation derived below: the individual vector charges are not separately conserved, whereas their sum is conserved. The conserved combination therefore characterizes the total singular content of the coupled rotational and frame-geometric sectors.

\section{Continuity equations}
\label{app:continuity_eqs}
Let us compute the continuity equations for our theory. We vary the source action of \eqref{eq.source_action_2} over the gauge fields, then
\be\label{eq:variation_sources_continuty}
\delta S_{\text{source}}  = \int dt d^2 x\,\left(\rho_k \delta\Phi_k + J_{ik} \delta A_{ik}  +   \tilde{j}_t \delta\tilde{a}_t +  \tilde{j}_j \delta\tilde{a}_j + j_t \delta a_t +  j_j \delta a_j  + \tilde{\rho}_k \delta \tilde{\Phi}_k + \tilde{J}_{ik} \delta\tilde{A}_{ik}  \right)\,,
\ee
and replacing the variation using the gauge transformation
\bea
\delta \tilde{\Phi}_k = \partial_t \tilde{\alpha}_k\,, \quad \delta \tilde{A}_{ik} = \partial_i \tilde{\alpha}_k\,, \\
\delta \Phi_k = \partial_t \alpha_k\,, \quad \delta A_{ik} = \partial_i \alpha_k\,, \\
\delta a_t = \partial_t \lambda\,, \quad \delta a_i = \partial_i \lambda + \epsilon_{ij}\alpha_j\,, \\
\delta \tilde{a}_t = \partial_t \lambda\,, \quad \delta \tilde{a}_i = \partial_i \lambda + \epsilon_{ij} \tilde{\alpha}_j\,,
\eea
these equations appear in eq. \eqref{eq:gauge_combined} in the main text. We can organize the terms in Eq. \eqref{eq:variation_sources_continuty}, and factor out the gauge functions $\lambda\,,  \alpha_i$ and $\tilde{\alpha}_i$. It may be necessary to integrate by parts. After all those steps, we can recognize the following relations
\bea\label{eq:continuty_eqs}
\partial_t \rho_k + \partial_i J_{ik} - \epsilon_{ik}j_i &=&0\,,  \nonumber \\
\partial_t( j_t + \tilde{j}_t) + \partial_i(j_i + \tilde{j}_i )  &=& 0 \,,  \\
\partial_t \tilde{\rho}_k + \partial_i\tilde{J}_{ik} - \epsilon_{ik} \tilde{j}_i &=& 0 \, \nonumber.
\eea
We observe that $j_i$ and $\tilde{j}_i$ act as source terms in the first and third continuity equations, respectively. Thus, the vector currents source
the corresponding tensor-charge sectors.

\section{Maxwell-like equations}
\label{app:maxwell_eqs}
In terms of magnetic and electric field, the HS fields can be write as 
\bea
P_k = B_k \,, &\quad& T_{ik} = \epsilon_{ij}\epsilon_{k\ell} E_{j\ell}\,. \\
\tilde{L}_t =  \tilde{b}\,, &\quad&  \tilde{L}_i = \epsilon_{ij}\tilde{e}_j\,, \\
L_t =  b\,, &\quad&  L_i = \epsilon_{ij}e_j\,, \\
\tilde{P}_k =  \tilde{b}_k\,, &\quad& \tilT_{ik} = \epsilon_{ij}\epsilon_{k\ell} \tilde{E}_{j\ell}\,.
\eea
Therefore, the dual lagrangian in eq. \eqref{eqs:HS_action_full} can write as
\bea\label{eq:elec_mag_full_lag}
\tilcalL &=& \tfrac{1}{2}\left(E_{ij}\tilde{C}_{ijkl} E_{kl} - B_i B_i \right)   - \tfrac{1}{2} ( \tilde{b} + A_{ii} )^2 + \tfrac{1}{2v^2} (\epsilon_{ij}\tilde{e}_j + \Phi_i )( \epsilon_{ik} \tilde{e}_k + \Phi_i ) \nonumber \\
&& - \tfrac{1}{2} ( b + A_{ii} )^2 + \tfrac{1}{2\beta} (\epsilon_{ij}e_j + \Phi_i )( \epsilon_{ik} e_k + \Phi_i ) \nonumber \\
&& - \tfrac{1}{2} \bigl[\tilde{B}_{k} - \epsilon_{ik} \bigl( \tilde{a}_i - a_i \bigr) \bigr] \bigl[ \tilde{B}_k - \epsilon_{\ell k} \bigl( \tilde{a}_\ell - a_\ell \bigr)) \bigr]  + \tfrac{1}{2} \tilde{K}_{ij k\ell} \bigl[ \tilde{E}_{k \ell}- \epsilon_{k \ell} \bigl( \tilde{a}_t - a_t \bigr) \bigr] \bigl[ \tilde{E}_{ij} - \epsilon_{ij} \bigl(  \tilde{a}_t - a_t  \bigr) \bigr] \,,
\eea
we used a compact notation for 
\be
\tilde{C}_{ijk\ell} = \epsilon_{ip} \epsilon_{jq} \epsilon_{km} \epsilon_{\ell n} \olC_{pqmn}\,, \quad  \tilde{K}_{ij k\ell} = \epsilon_{ip} \epsilon_{jq} \epsilon_{km} \epsilon_{\ell n} \olK_{pqmn} \,,
\ee
these tensor constants have to follow the same symmetries as eq. \eqref{eq:symmetries_C_K} above.
Now, let us rewrite the last the set of Maxwell-like equations for each sector in our model.

In case of the tensor gauge theory $(\Phi_k, A_{ik})$ thus
\bea
\overline{C}_{ikj\ell}  \epsilon_{im}\epsilon_{jp}\epsilon_{\ell q} \partial_m E_{pq} &=& -\rho_k\,, \quad \text{(Gauss Law)} \\
\partial_t B_k - \epsilon_{ij} \epsilon_{k\ell} \partial_i E_{j\ell}&=& 0 \,, \quad \text{(Faraday Law)}\\
\epsilon_{im} \left[\partial_i B_k - \overline{C}_{ikj\ell}  \epsilon_{jp} \epsilon_{\ell q}\partial_t E_{pq}\right] &=& -J_{mk}\,. \quad \text{(Amp\'ere Law)}
\eea
For one of the $U(1)$ field given by $(a_t, a_i)$ are
\bea
\frac{1}{\beta} \epsilon_{ij}\partial_j (\epsilon_{ik}e_k +\Phi_i)  &=& -j_t\,, \quad \text{(Gauss Law)} \\
\partial_t b- \epsilon_{ij}  \partial_i e_j &=& 0 \,, \quad \text{(Faraday Law)}\\
-\frac{1}{\beta}\epsilon_{ij}\partial_t(\epsilon_{ik}e_k + \Phi_i)  +\epsilon_{ij}\partial_i (b + A_{kk}) &=& - j_j \,,  \quad \text{(Amp\'ere Law)}
\eea
In case of the second set of $U(1)$ field describe by $(\tilde{a}_t, \tilde{a}_i)$ are
\bea
\frac{1}{v^2} \epsilon_{ij}\partial_j (\epsilon_{ik} \tilde{e}_k +\Phi_i)  &=& -\tilde{j}_t\,, \quad \text{(Gauss Law)} \\
\partial_t \tilde{b} - \epsilon_{ij}  \partial_i \tilde{e}_j &=& 0 \,, \quad \text{(Faraday Law)}\\
-\frac{1}{v^2}\epsilon_{ij}\partial_t(\epsilon_{ik}\tilde{e}_k + \Phi_i)  +\epsilon_{ij}\partial_i (\tilde{b} + A_{kk}) &=& - \tilde{j}_j \,,  \quad \text{(Amp\'ere Law)}
\eea
The second tensor gauge field $(\tilde{\Phi}_k, \tilde{A}_{ik})$ are
\bea
\overline{K}_{ijk \ell}\epsilon_{im} \partial_m \left[ \epsilon_{kp}\epsilon_{\ell q} \tilde{E}_{pq}  - \epsilon_{k\ell}(\tilde{a}_t - a_t) \right] &=& - 	 \tilde{\rho}_j\,,  \quad \text{(Gauss Law)}\\
\partial_t \tilde{B}_k - \epsilon_{ij} \epsilon_{k\ell} \partial_i \tilde{E}_{j\ell}&=& 0 \,, \quad \text{(Faraday Law)}\\
\epsilon_{im} \partial_i \left[ \tilde{B}_\ell  - \epsilon_{ k\ell}(\tilde{a}_k - a_k )\right] - \overline{K}_{ijk\ell} \epsilon_{km} \partial_t \left[ \epsilon_{ip}\epsilon_{jq}\tilde{E}_{pq} - \epsilon_{ij} (\tilde{a}_t - a_t)\right]  &=& - \tilde{J}_{m\ell}\,. \quad \text{(Amp\'ere Law)}
\eea

\section{Fractonic behavior}
\label{app:fractonic_behavior} 

In this appendix we derive an integrability condition on the composite current which underlies the mobility constraints
discussed in the main text. We use the continuity equations given in eqs. \eqref{eq:continuty_eqs}. We add the two tensor continuity equations
\be
\partial_t \rho_k + \partial_i J_{ik} - \epsilon_{ik} j_i = 0 \,,\quad
\partial_t \tilde{\rho}_k + \partial_i \tilde{J}_{ik} - \epsilon_{ik} \tilde{j}_i = 0\,,
\ee
which yields
\be
\partial_t(\rho_k+\tilde{\rho}_k) +\partial_i(J_{ik}+\tilde{J}_{ik}) -\epsilon_{ik}(j_i+\tilde{j}_i)=0 \,.
\ee
Next, we contract this equation with the Levi-Civita tensor $\epsilon_{\ell k}$, obtaining
\be
\partial_t\!\left[\epsilon_{\ell k}(\rho_k+\tilde{\rho}_k)\right]
+\partial_i\!\left[\epsilon_{\ell k}(J_{ik}+\tilde{J}_{ik})\right]
-(j_{\ell} +\tilde{j}_{\ell})=0 \,,
\ee
where we used $\epsilon_{\ell k}\epsilon_{ik}=\delta_{\ell i}$. Taking a spatial
divergence of this equation with $\partial_{\ell}$ gives
\be
\partial_t\!\left[\epsilon_{\ell k}\partial_{\ell} (\rho_k+\tilde{\rho}_k)\right]
+\partial_i\!\left[\epsilon_{\ell k}\partial_{\ell} (J_{ik}+\tilde{J}_{ik})\right]
-\partial_{\ell} (j_{\ell}+\tilde{j}_{\ell})=0 \,.
\ee
Using the $U(1)$ continuity equation
$\partial_t(j_t+\tilde j_t)+\partial_i(j_i+\tilde j_i)=0$, this becomes
\begin{equation}
\partial_t\!\left[\epsilon_{\ell k}\partial_{\ell} (\rho_k+\tilde{\rho}_k)
+(j_t+\tilde j_t)\right]
+\partial_i\partial_{\ell} \!\left[\epsilon_{\ell k}(J_{ik}+\tilde{J}_{ik})\right]=0 \,.
\end{equation}
We rewrite the continuity equation as
\begin{equation}
\partial_t \rho_{\mathrm{tot}} + \partial_i\partial_j\,\mathcal{J}_{ij}=0 \,,
\label{eq:continuity_general_eq}
\end{equation}
where we have defined the composite density and two-index current
\begin{equation}
\rho_{\mathrm{tot}} \equiv \epsilon_{ij}\partial_i (\rho_j+\tilde{\rho}_j) + (j_t+\tilde j_t),
\qquad 
\mathcal{J}_{ij} \equiv \epsilon_{jk}\big(J_{ik}+\tilde J_{ik}\big).
\label{Eq:rhotot}
\end{equation}
We can define the dipole and the trace of the quadrupole moments
\begin{equation}
D_k = \int d^2x\, x_k \rho_{\mathrm{tot}}, 
\qquad 
Q_{kk} = \int d^2x\, x^2 \rho_{\mathrm{tot}} .
\end{equation}
Taking a time derivative of the dipole moment gives
\begin{equation}
\frac{dD_k}{dt}
= \int d^2x\, x_k \partial_t \rho_{\mathrm{tot}}
= - \int d^2x\, x_k \partial_i\partial_j \mathcal{J}_{ij}
= \int d^2x\, \partial_j \mathcal{J}_{kj} ,
\label{eq:dipole}
\end{equation}
where in the last step we integrated by parts and neglected boundary terms.

Assuming the absence of boundary fluxes, the spatial integral in
Eq.~\eqref{eq:dipole} must vanish for arbitrary integration domains. Therefore, the integrand must vanish locally, thereby implying the local integrability condition
\begin{equation}
\partial_j \mathcal{J}_{ij}
=
\partial_j\!\left[\epsilon_{jk}\big(J_{ik}+\tilde J_{ik}\big)\right]
=
\epsilon_{jk}\,\partial_j\big(J_{ik}+\tilde J_{ik}\big)
=0 .
\end{equation}
This relation expresses a kinematic constraint on the composite current
$\mathcal{J}_{ij}$ defined in Eq.~\eqref{Eq:rhotot}. In terms of the
singular fields, it corresponds to a compatibility condition on defect
configurations generated by the displacement and frame fields.
Consequently, the mobility restrictions discussed in the main text, such as the glide constraint, follow directly from this integrability condition together with the
defect continuity equations.

\end{widetext}

\bibliography{main}

\end{document}